\documentstyle{article}

\input epsf

\begin{document}

\title{Reactions at polymer interfaces: A Monte Carlo Simulation} 

\author{M.\ M\"{u}ller${}^{1,2}$ 
\\
{\small ${}^1$ Department of Physics, Box 351560, University of Washington} \\
{\small Seattle, Washington 98195-1560}\\
{\small and}\\
{\small ${}^2$ permanent address: Institut f{\"u}r Physik, Johannes Gutenberg Universit{\"a}t} \\
{\small D-55099 Mainz, Germany}}
\date{\today}
\maketitle

\begin{abstract}
Reactions at a strongly segregated interface of a symmetric binary
polymer blend are investigated via Monte Carlo simulations.  End
functionalized homopolymers of different species interact at the
interface instantaneously and irreversibly to form diblock
copolymers. The simulations, in the framework of the bond fluctuation
model, determine the time dependence of the copolymer production in the
initial and intermediate time regime for small reactant concentration
$\rho_0 R_g^3=0.163 \cdots 0.0406$. The results are compared to recent
theories and simulation data of a simple reaction diffusion model. For
the reactant concentration accessible in the simulation, no linear
growth of the copolymer density is found in the initial regime, and a
$\sqrt{t}$-law is observed in the intermediate stage.

\end{abstract}

\section{ Introduction }
Reactions at interfaces have attracted great interest because of a
variety of practical applications.  They are involved in facilitated
transport across biological membranes, and in selective removal of a
component of a mixture by contact with a reactive
liquid\cite{REACL}. Recent interest stems from extensive commercial
application of reactive polymer blending technique\cite{EX1,EX3}
and progress of the experimental studies\cite{EX4,EX5,EX6}.
Melt blending of different polymer species (denoted A and B) is widely
employed\cite{APPL} to improve the application properties of polymeric
materials. Since the entropy of mixing of high-molecular-weight polymers
is extremely low, macromolecular alloys can often be described as an
assembly of interfaces between relatively pure phases. Block copolymers
containing both types of monomers can be used to tailor the properties
of polymer-polymer interfaces. They promote mixing by means of reducing the
interfacial tension and the droplet-droplet coalescence rate\cite{EX4}
and mechanically strengthening the composite material due to entanglements
across the interface. An effective method of creating copolymers at the
interface is {\em in-situ} production\cite{EX1}, e.g.\ chemical
reactions of end functionalized homopolymers.

Recently, reactions at a flat interface in a binary, symmetric polymer
blend have been studied independently and in some detail by O'Shaughnessy
and Sawhney\cite{S1,S2} and Fredrickson\cite{F1} and Milner\cite{F2}: A
and B polymers are assumed to be immiscible, but structurally
symmetric. The almost pure phases A and B host a small fraction $\rho_0$
of chains with a terminal reactive group. The A reactive ends can only
react with B reactive ends and form AB diblock-copolymers at the
interface between the bulk phases. The reactions are instantaneous and
irreversible.  The interfacial density $\sigma$ of copolymers at the
interface grows according to the rate equation:
\begin{equation}
\frac{d\sigma}{dt} = K(t) \rho_0^2
\end{equation}
where $t$ measures the duration of the reaction and $\rho_0$ denotes
the time independent fraction of reactive chains in the bulk.
The reaction rate $K$ is largely independent from the details of the
microscopic mechanism, because subdiffusive monomer motion on small time
scales guarantees reaction whenever the coils of two reactive chains
overlap\cite{BDYN}. For small number density $\rho_0$ of reactive chains
($\rho_0 R_g^3 \ll 1$ with $R_g$ radius of gyration) and high
incompatibility, Fredrickson and Milner\cite{F2} predict three time
regimes: At times smaller than $\tau=D/K_0^2\rho_0^2$, the density of
reactive chains at the interface remains close to its bulk value. The
reaction rate within the framework of the Rouse model\cite{ROUSE} is
time-independent, and
given by\cite{S1,F1}: $K_0 \approx 100.6 D R_g^2/\ln N$. It follows that
the interfacial
density of copolymers grows linearly with time in this regime. At times larger
than $\tau$\cite{C1}, there is a depletion hole of reactive chains with spatial
extension $\sqrt{Dt}$, and the reaction rate is determined by the flux of
reactive chains to the interface. The reaction rate decays as
$K=\sqrt{D/\pi\rho_0^2t}$, {\em i.e.} the interface acts much like an
absorbing boundary. The copolymer density grows like $t^{1/2}$ in this
regime. For even longer times $t > \Phi^2
R_g^2/N^2D\rho_0^2$ ($\Phi$ monomer density), the copolymer accumulates at 
the interface and forms a brush. This prevents  reactive chains from reaching the
interface, and the reaction rate becomes extremely small, decaying
as $1/t \sqrt{\ln t}$. The copolymer density increases only as $\sqrt{\ln t}.$

In spite of the practical applications of interfacial reactions,
qualitative experiments to determine the time dependence of the
copolymer concentration at the interface are still
rare\cite{EX1,EX3,EX4,EX2}. Their interpretation is complicated by
structural asymmetries between the two homopolymer species, or the
reactive and inert polymers of the same species.  Furthermore, large
differences in  the mobilities of the species lead to qualitative
modifications of the reaction kinetics\cite{IDYN}.  In such a situation,
Monte
Carlo simulations can provide valuable information for understanding 
the reactions in
these inhomogeneous, complex fluids and for exploring the validity of
approximations in the analytic treatment. Such
simulations of reactions at surfaces and of catalysis of small molecules
have proven extremely useful in studying the stochastic dynamics of
reaction-diffusion processes\cite{Albano}.  The present simulations are
the first to investigate reactions at polymer interfaces, and they cover the
initial and intermediate time regime. They yield the time evolution, as
well as the spatial depletion profile, of reactants at the interface. We
compare our results to the theories described above and to a simple
reaction-diffusion model.

\section{ Model and simulation technique }
If the microscopic reaction is sufficiently fast, the reaction rate is
largely independent of the microscopic mechanisms\cite{S1,BDYN}. Since
the study of reaction kinetics poses high computational demands, we
employ a coarse grained and highly computational efficient lattice model
of a dense polymeric melt. The Bond Fluctuation Model (BFM)\cite{BFM}
retains the relevant features of polymeric materials: connectivity of
the monomers along a chain molecule, excluded volume interaction of the
monomeric units, and a short range thermal interaction potential.  The
BFM has proven very useful for studying various related properties of
dense polymeric melts. Among them are single chain dynamics in athermal
melts\cite{PAUL,NOTROUSE} and binary blends\cite{DYN}, living
polymerization\cite{LIVE}, adsorption in polymer brushes\cite{ACHIM},
and interfacial properties in binary\cite{INTER,INTER2} and ternary
blends\cite{TERNARY}.  In the framework of the BFM each segment occupies
all eight sites at the corners of a unit cell in a simple cubic lattice,
and no site can be doubly occupied. Adjacent monomers along a chain
molecule are connected via one of 108 bond vectors. These are chosen
such that the local self-avoidance prevents chains from crossing each
other during their motion. The fluctuating bond length permits the
implementation of random, local monomer displacements\cite{BFM} such
that the dynamics of an {\em isolated} chain is well describable by the
Rouse model\cite{PAUL,NOTROUSE}. We work at a filling fraction
$8\Phi=0.5$ of occupied lattice sites, and all length are measured in
units of the lattice spacing $u$. At this density the BFM reproduces many
characteristics of a dense polymer melt.  The blend consists of two
species of homopolymers, denoted A and B, of equal chain length $N=32$,
and diblock copolymers AB of chain length 64. This is well below the
chain length $N=200$, where first signs of repation-like motion are
observed. We employ a square well potential, which encompasses the first
peak of the pair correlation function, i.e.\ comprising all 54 neighbor
sites up to a distance $\sqrt{6}$ lattice units:
$\epsilon=\epsilon_{AB}=-\epsilon_{AA}=-\epsilon_{BB}=0.1
k_BT$. Monomers of the same kind attract each other, whereas there is a
repulsion between unlike species. This incompatibility corresponds to
$\chi N \approx 17$, and is well inside the strong segregation limit for
binary blends. Both, the interfacial properties of the binary
blend\cite{INTER,INTER2} as well as the equilibrium behavior of copolymers at
interfaces\cite{TERNARY,WERNER} have been investigated extensively.

We work in a $64 \times 64 \times 128$ geometry, and for the highest
fraction of reactive chains additional data were obtained for system
size $64 \times 64 \times 256$. Periodic boundary conditions in all
three spatial directions were applied, such that the systems contain two
interfaces parallel to the xy-plane. The initial configurations for the
binary blend have been carefully equilibrated using $10^6$ local
MCS. (On the average each monomer attempts to jump once within a Monte
Carlo Step/MCS). After this initial equilibration of the interfacial
properties, the equilibration was continued and starting configurations
for the reaction study were collected every 50,000 MCS. For the smaller
system size we average our results over 96 independent simulation runs,
corresponding to a total of more than 3 million monomers. For the larger
size we employ 30 independent configurations. Simulations were performed
on a CONVEX SPP1200, using a trivial parallelization strategy.

\begin{figure}[htbp]
    \begin{minipage}[t]{160mm}%
       \setlength{\epsfxsize}{13cm}
       \mbox{\epsffile{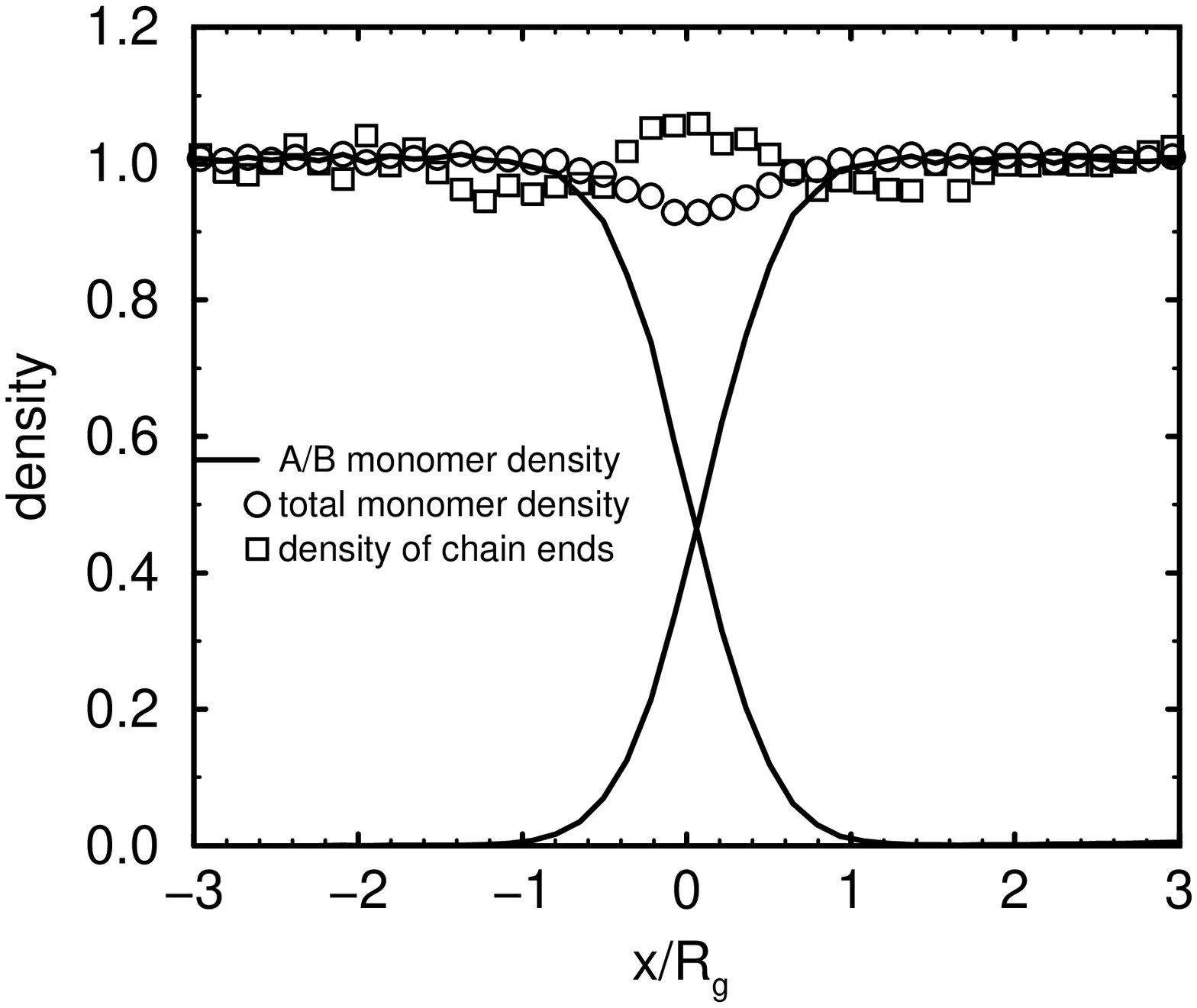}}
    \end{minipage}%
    \hfill%
    \begin{minipage}[b]{160mm}%
       \caption{Interfacial properties of the equilibrated starting configuration: The solid lines present the
                normed density of A and B monomers, wheras the symbols denote the total monomer density (circles) and  the
                density of chain ends (squares) scaled by a factor $N/2$. Note the enrichment of chain ends at the center
                of the interface.
                }
       \label{fig:prof1}
    \end{minipage}%
\end{figure}

Some interfacial properties of the starting configurations are presented
in Fig. \ref{fig:prof1}.  The blend is well segregated, i.e.\ the
coexisting phases consist almost entirely of A or B polymers.  Between
the coexisting phases there is an interface of width $w \approx 3.7$.
The conformational properties have been discussed
in detail in Ref.\cite{INTER,INTER2}. We simply emphasize that, in agreement with
mean field theories\cite{INTER2}, there is a pronounced segregation of
chain ends at the interface and a concomitant depletion at a distance
$R_g$. The total monomer density is somewhat reduced at the
center of the interface.

In these {\em equilibrated} starting configurations a fraction of
homopolymers was randomly chosen to bear a reactive end, and the
reaction was started at time $t=0$. This corresponds to an experimental
situation in which a (fast diffusive) reactive agent/catalyst\cite{EX1} is
added to the blend, or the reaction is initiated by radiation.
O'Shaughnessy and Sawhney\cite{S1,S2} analyzed the qualitative dependence 
on the monomeric reaction probability. For very low monomeric reaction 
probability $Q$, correlation in the density of reactive particles are
only weakly perturbed by the reactions, and the reaction rate $K$ is 
proportional to the equilibrium contact probability of reactive ends.
However, this mean field kinetics is only appropriate if the reaction
probability is smaller than a crossover value $Q^{\star} \sim 1/\ln N$\cite{S1,S2}.
Therefore, reactions in high molecular weight polymers are typically 
diffusion controlled. This diffusion controlled reaction kinetics is largely 
independent from the details of the monomeric reaction mechanism, especially 
from the monomeric capture radius and reaction probabilities. Since the
simulation can treat only rather short polymers, we use a large
monomeric reaction rate $Q$ and capture radius to make sure that we
work in the controlled regime.
If the distance between reactive A-ends and B-ends is less
than the capture radius $\sqrt{6}$, reactive homopolymers form
AB diblock copolymers instantaneously and irreversibly. 
Of course, the concrete value of the capture radius $\sqrt{6}$
has no physical significance; it is the largest value the model
allows to be compatible with local monomer dynamics and chain 
connectivity.
According to analytical treatments\cite{S1,S2,F1,F2}, we 
expect our results to be characteristic for diffusion controlled
kinetics and rather independent of the concrete monomeric reaction 
mechanism (e.g. the model also applies to long polymers with a large
but finite monomeric reaction rate).

Otherwise the reactive
chains behave the same as the non-reactive polymers. We considered systems
with
three different values of the initial number density of reactive chains 
in their corresponding
bulk phases: $\rho_0 = 1/2048, 1/4096$, and $1/8192$. With the
measured radius of gyration $R_g=6.93$, the scaled number density ranged
in our simulation from $\rho_0 R_g^3=0.041$ to $0.1625$. Furthermore, we
determined the self-diffusion constant from the mean square
displacements in the plane parallel to the interface for the largest
system size to $D=1.17(5) \cdot 10^{-4} u^2/\mbox{MCS}$, which is very close to the
corresponding athermal value.

\section{ Simulation results }
\begin{figure}[htbp]
    \begin{minipage}[t]{160mm}%
       \setlength{\epsfxsize}{13cm}
       \mbox{\epsffile{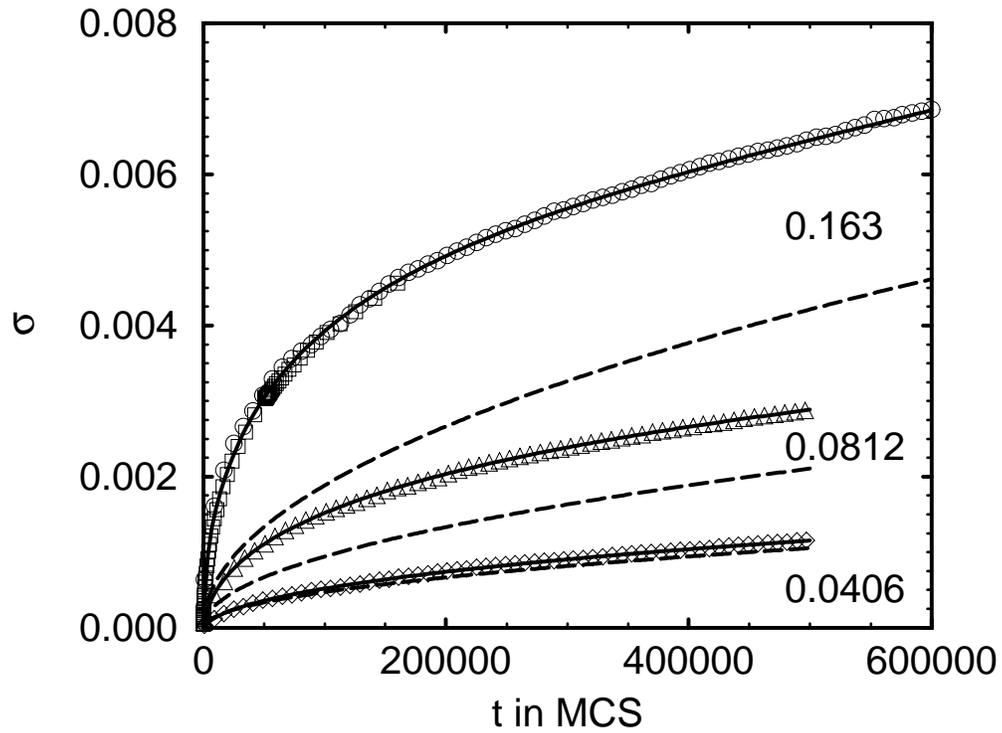}}
    \end{minipage}%
    \hfill%
    \begin{minipage}[b]{160mm}%
       \caption{Time dependence of the copolymer concentration at the interface: 
                circles denote simulation data for $\rho_0=1/2048$ and $64 \times 64 \times 256$ system size,
                squares correspond to $\rho_0=1/2048$, triangles to $\rho_0=1/4096$, diamonds to $\rho_0=1/8192$
                and system size $64 \times 64 \times 128$ respectively. The solid lines are fits to the simulation raw
                data, whereas the dashed lines represent the predicted time development in the intermediate regime.
                }
       \label{fig:time}
    \end{minipage}%
\end{figure}

Upon activating the reactive polymers, AB diblock copolymers are created
at the interface. The time dependence of the interfacial density for
different fraction of reactive chains is presented in
Fig. \ref{fig:time}. For the highest fraction, the simulation data for
the two system sizes agree to within the statistical accuracy. Since
finite size effects are expected to be even smaller for lower fractions
of reactive chains, they can be neglected for our data.  In contrast to
the predictions above, the simulation data for the copolymer interfacial
density do not exhibit a linear growth law at small times. Due to the
high initial reaction rate in the Monte Carlo simulation, the theory
underestimates the interfacial density over the entire time covered by
the simulation for the two largest reactant concentrations. The broken
lines in the figure show the theoretical prediction for the diffusive
growth in the intermediate time regime. For the lowest density, the
simulation data are rather well described by the prediction for the
intermediate regime. However the agreement is fortuitous, as the
simulation data cover mainly the initial time regime ($\tau \approx 3
\cdot 10^5$ MCS).

\begin{figure}[htbp]
    \begin{minipage}[t]{160mm}%
       \setlength{\epsfxsize}{13cm}
       \mbox{\epsffile{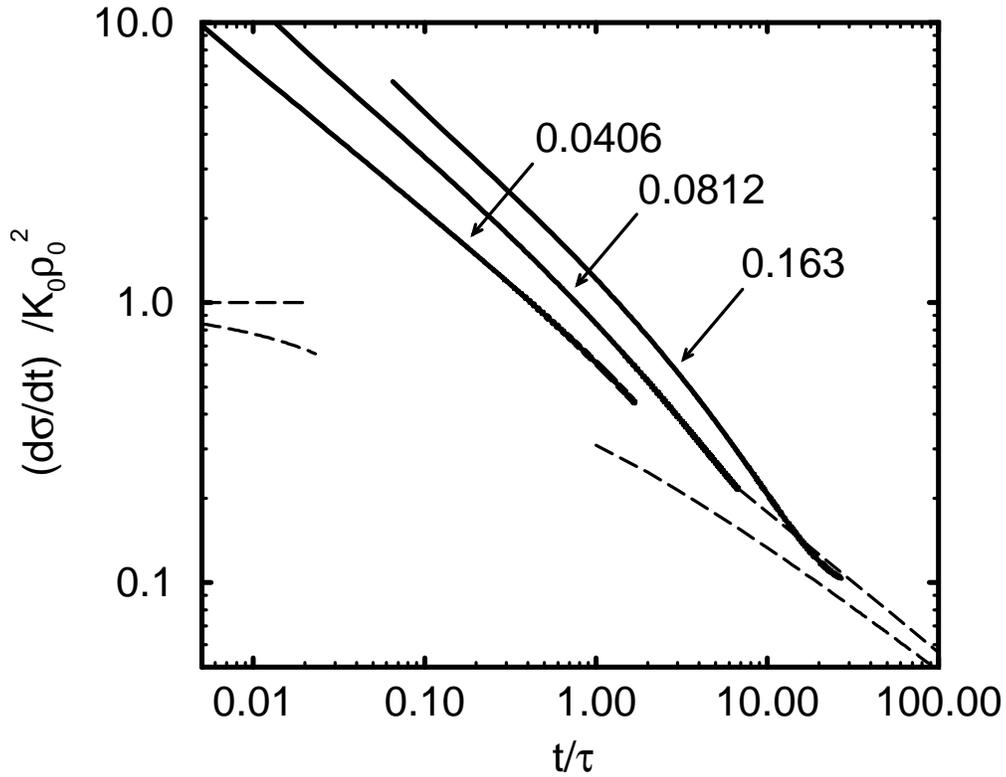}}
    \end{minipage}%
    \hfill%
    \begin{minipage}[b]{160mm}%
       \caption{Scaled reaction rate $\frac{d\sigma/dt}{K_0\rho_0^2}$ for the polymer system: 
                The solid lines are the results of the Monte Carlo simulation for different
                values $\rho_0 R_g^3$ as indicated in the figure. The two set of dashed lines 
		correspond to the predictions of Fredrickson and Milner[11] for $\rho_0 R_g^3 \ll 1$.
		The higher values correspond to the asymptotic results for small and large values of $t/\tau$;
		the lower ones present first order corrections.
                }
       \label{fig:k}
    \end{minipage}%
\end{figure}

To eliminate the influence of the fast initial rate, it is instructive
to monitor the growth rate $K(t)$.  The solid lines in
Fig. \ref{fig:time} show fits to the time development of the
copolymer density which are used to determine the reaction rate. These
rates are presented in Fig. \ref{fig:k} in scaled form.  The solid lines
show the simulational results, while the dashed ones correspond to
the theoretical predictions in the initial ($t<\tau$) and intermediate
($t>\tau$) regimes. First order corrections\cite{F2} to describe the
crossover between the two regimes are also included into the figure:
\begin{eqnarray}
\frac{d\sigma/dt}{K_0 \rho_o^2} &=& 1 - \frac{4}{\sqrt{\pi}}
\left(\frac{t}{\tau}\right)^{1/2} \qquad \mbox{\hspace*{2.3cm}for}
\qquad t \ll \tau \qquad \mbox{initial regime}\\ \frac{d\sigma/dt}{K_0
\rho_o^2} &=& \frac{1}{\sqrt{\pi}} \left(\frac{t}{\tau}\right)^{-1/2} -
0.254 \left(\frac{t}{\tau}\right)^{-3/4}
\qquad \mbox{for} \qquad t \gg \tau \qquad \mbox{intermediate regime}
\end{eqnarray}
 Even for $\rho_0 R_g^3$ as small as $0.0406$ the prediction
underestimates the reaction rate by an order of magnitude, and there is
{\em no} sign of a constant reaction rate at small times. Furthermore
the simulation data for different fractions of reactive chains do not
collapse onto a single curve. In the intermediate time regime, however, 
the simulation data are consistent with the theoretical prediction $K
\sim 1/\sqrt{t}$, and the anticipated proportionality constant is in rough
agreement with the simulation data, though the asymptotic limit $t/\tau \gg 1$
is not reached for the lower fractions of reactive chains.

\begin{figure}[htbp]
    \begin{minipage}[t]{160mm}%
       \setlength{\epsfxsize}{13cm}
       \mbox{\epsffile{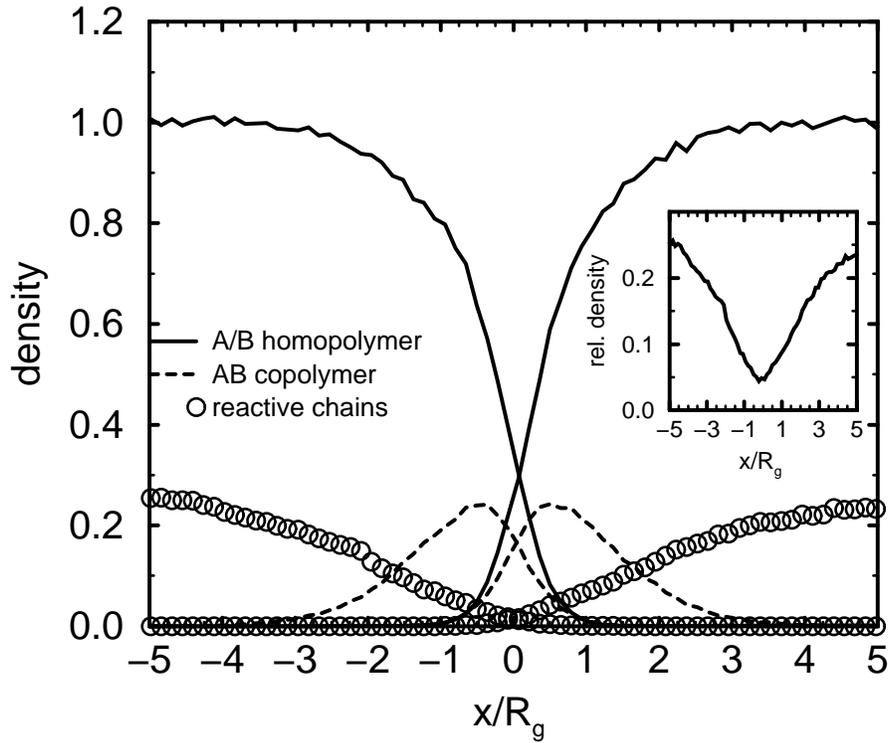}}
    \end{minipage}%
    \hfill%
    \begin{minipage}[b]{160mm}%
       \caption{Non-equilibrium density profiles for $\rho_0 R_g^3=0.163$ and $t/\tau=24.4$. The solid lines present the 
                monomer density
                of the A and B segments of the homopolymers. The dashed lines correspond to the A and B monomers of the 
                copolymers, whereas the monomer density of the two types of reactive chains is denoted by circles.
                The inset shows the ratio of the monomer density of reactive chains to the total monomer density of
                homopolymers.
                }
       \label{fig:p60}
    \end{minipage}%
\end{figure}

In Fig.\ \ref{fig:p60} we present the non-equilibrium profiles across
the interface for $\rho_o R_g^3=0.163$ and $t/\tau\approx 24.4$.  The
copolymers accumulate at the interface, and their composition profile 
resembles its equilibrium shape, as determined
in Ref.\ \cite{WERNER}. The copolymers do not yet form a brush at the
interface. Therefore the simulation data are still in the intermediate
regime. The homopolymer density profile, however, is altered as they are
displaced from the interface. The simulation data
also confirm the predicted depletion of reactive chains in the
interfacial zone, and the growth of this depletion hole with time.  Thus the
simulation data corroborate the main features of the theory in the
intermediate time regime, but disagree strongly with predictions 
in the initial stage.

There are several possible reasons for the difference in the initial stage 
between the simulation data and the theory:

The local equilibrium structure of the homopolymer interface exhibits an
enrichment of chain ends at the interface. Therefore the initial reaction rate
is expected to be higher than that for a uniform initial distribution of reactive 
segments. This effect has been used to rationalize the enhanced reaction rate in
PS-PMMA interfaces\cite{EX2}.  However, it is not clear to what extent this 
correlation in the initial condition accounts quantitatively for the discrepancies 
between simulation and theory.
Futhermore, the theory deals with Rouse chains at an interface in the Helfand regime
$1 \gg \chi \gg 1/N$. This latter condition can be satisfied only for very long chains, 
whereas the chain length $N$ must not be too large so that the chain dynamics is 
describable by the Rouse model.

Additionally, the reactant density $\rho_o R_g^3=0.0406$ in the simulations might still not be 
small enough to meet the condition $\rho_o R_g^3 \ll 1$ assumed in the theory.  
The constant reaction rate should be observable for times much larger than
the relaxation time $\tau_R$  of the internal chain motion, but much smaller
than the crossover time $\tau$ to the intermediate regime. Note however, that 
the crossover time scale $\tau$ and the Rouse time $\tau_R={6R_g^2}/{\pi^2D} 
\approx \tau (78.6 \rho_0 R_g^3/\ln N)^2$ are not well separated for the reactant 
concentration studied. Unfortunately, simulation at even lower fraction of 
reactive chains pose very high statistical demands, which are beyond our 
computational facilities.

\section{ Reaction diffusion model }
We decided to compare our results to additional simulations of a simple 
random walk model.  Bulk reactions\cite{BDYN} in polymeric systems can be 
modeled schematically by random walkers which react if their mutual 
distance is smaller than the radius of gyration in the corresponding 
polymer system.  Therefore we consider two types of random walkers, denoted 
A and B, on a simple cubic lattice.  A-type walkers are restricted to the 
half-space $z\geq 0$, whereas B-type walkers are confined to $z\leq 0$.  An 
A-type walker annihilates with a B-type walker (and vice versa), whenever 
their mutual distance is smaller than $R$.  The initial time regime for a 
class of similar models has been discussed by Durning and O'Shaugnessy 
\cite{IDYN}.  The model is an appropriate caricature of polymer reactions 
at interfaces for the initial and intermediate time regime.  Since it 
eliminates the shortest time scales, which are associated with the 
segmental motion on distances smaller than the radius of gyration, and also 
neglects the inert homopolymer matrix, much lower volume fractions for 
reactive particles are accessible in the simulations.  Furthermore, many of 
the above-mentioned complications characteristic of polymeric systems are 
absent.  We work with a capture radius $R=8$ and a $256 \times 256 \times 
512$ geometry, and a $1024 \times 1024 \times 2048$ system for the lowest density.
Initially, the random walkers are uniformly distributed in 
their volumes.  We employ at least 512 independent systems.

\begin{figure}[htbp]
    \begin{minipage}[t]{160mm}%
       \setlength{\epsfxsize}{13cm}
       \mbox{\epsffile{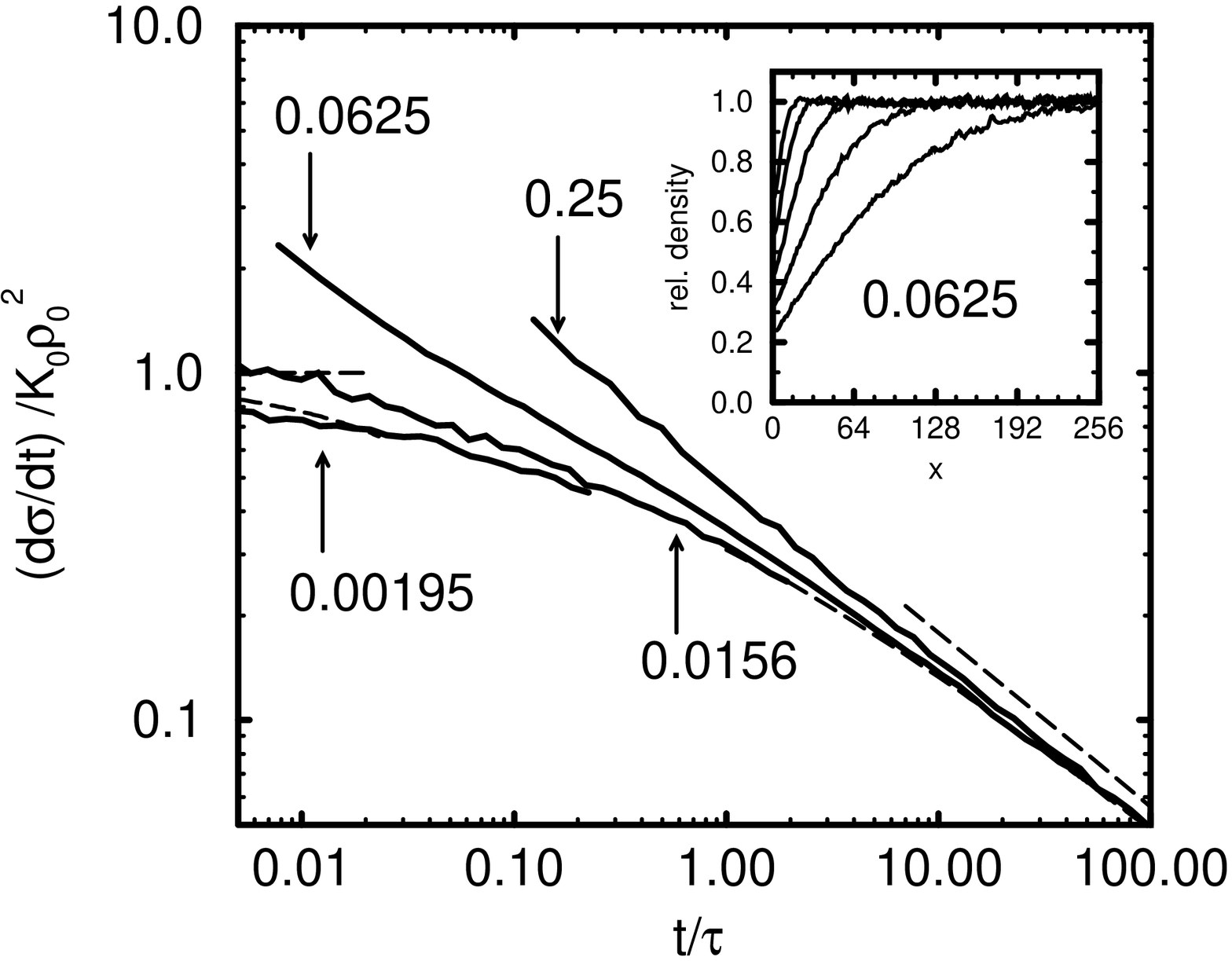}}
    \end{minipage}%
    \hfill%
    \begin{minipage}[b]{160mm}%
       \caption{Scaled reaction rate for the reaction diffusion system:
                The solid lines are the result of the Monte Carlo 
                simulations, the values of $\rho_0 R_g^3$ being
                indicated in the figure. The dashed lines are the same as 
                in Fig.3. Only for very small concentration does
                one observe a reaction rate which is in agreement with the predictions [9]. The initial reaction rate
                is $K_0\approx 157$ \newline
                The inset presents the density profiles of reactants for $\rho_0 R_g^3=0.0625$ and $t/\tau=0.28, 1.13,
                4.53, 18.12$ and $72.49$.
                }
       \label{fig:kdr}
    \end{minipage}%
\end{figure}

The reaction rate in its scaled form as well as the prediction of 
Fredrickson and Milner\cite{F2} are presented in Fig.  \ref{fig:kdr}.  For very 
small density of walkers $\rho_0 R^3<0.002$, the simulations show an almost 
constant initial reaction rate, in agreement with the predictions.  The 
crossover to the intermediate regime, and the diffusive growth in the 
intermediate regime, is described by the theory in the limit of vanishing 
reactant concentration, and the corresponding scaling function is obtained 
as the lower envelope of the simulation data.  The spatial density profile 
(cf.\ Fig.\ \ref{fig:kdr} inset) in the intermediate time regime exhibits a 
pronounced depletion hole, and is qualitatively similar to the profile 
obtained in the simulation of the polymer system.  In the vicinity of the 
interface there are some deviations due to the finite interfacial width in 
the polymer system.

For concentrations comparable to those used in the simulation of the polymer systems, 
the reaction diffusion model shows similar deviations, i.e.\ a decaying reaction rate in the 
initial stage which is always larger than in the limit $\rho_0 \to 0$. However the 
deviations are somewhat smaller than for the polymer system. This is partially due to 
the fact, that the reaction in the polymer system does not take place instantaneous when the chain
volumes of the reactive pair overlap, but the reaction is retarded by a time of the order $\tau_R$.
Therefore we conclude that the finite concentration of reactive chains accounts qualitatively for 
the discrepancies between the analytic theory and simulations of the polymeric system. However, 
the simple reaction-diffusion model cannot describe the polymeric system quantitatively for finite 
reactant concentrations.

\section{ Conclusions and outlook} 
In summary, we have presented extensive 
simulations of reactions at interfaces, both for polymeric systems as well 
as for a simple reaction-diffusion model.  We have compared our simulation 
results to recent theoretical predictions\cite{S1,S2,F1,F2}.  At 
intermediate times we find agreement between the theory and the simulations 
of both models.  The initial state and the crossover to the intermediate 
regime is described by the theory {\em only for extremely small 
concentration of reactive particles}.  These are not accessible in the 
simulation of the polymeric systems, but are in the simulation of the 
random walker model.  Results of the latter are compatible with the 
theoretical predictions in the limit $\rho_0 R_g^3 \to 0$, and the 
simulations yield estimates for the complete crossover scaling behavior of 
the reaction rate.  However, for non-zero but small concentrations of 
reactive particles, the reaction rate is grossly underestimated by the 
theory.  Even for the smallest fraction of reactive chains simulated in the 
present study, $\rho_0 R_g^3 = 0.0406$ , the initial reaction rate obtained 
from the simulation is about an order of magnitude larger than predicted, 
and is not constant, but decays with time.  
It should be emphasized that 
our results are not an artifact of the small chain length used in the 
simulation.  The reactant concentrations correspond to weight fractions of 
$25\%, 12.5\%$ and $6.25\%$.  These weight fractions are within the 
experimentally relevant range.  If one increases the chain length at fixed 
weight fraction, then $\rho_0 R_g^3$ increases as $\sqrt{N}$ taking the 
initial condition even further from the assumptions of the theory.  Therefore we 
expect that experiments will also observe a decay of the reaction rate over 
the entire time regime, unless extremely small volume fractions of reactive 
chains are used. 

 Further studies might investigate the crossover between 
the diffusive growth and the late regime, in which the copolymer brush 
hinders the reactive chains to reach the surface.  This regime is 
accessible for high concentration of reactive chains.  At high 
incompatibilities the copolymers cannot reduce the surface tension 
completely, and there is a first order transition to a lamellar phase at 
high copolymer concentration\cite{TERNARY}.  The direct measurement of the 
interactions between copolymer ``coated'' interfaces\cite{TERNARY} is an 
interesting topic, because it relates to the important effect of a reduced 
droplet-droplet coalescence rate.  In the weak segregation limit, however, 
the addition of copolymer causes a continuous transition to the disordered 
phase to be encountered, (the Scott-line which exists for $2<\chi N<3$ 
within mean field theory\cite{MF}).  Thus it would be interesting to study 
the temperature dependence of the reaction kinetics at weak and 
intermediate segregations, where reactions might lead to complete 
miscibility.
\subsection*{Acknowledgement}
It is a great pleasure to thank M.\ Schick for valuable discussions and
critical reading of the manuscript.
Financial support by the Alexander von Humboldt foundation and the 
National Science Foundation under Grant No.\ DMR9531161 as well as
a generous allocation of computer time on the CONVEX SPP 1200 of the
computing center in Mainz, Germany, are gratefully acknowledged.

\begin{thebibliography}{99}
\bibitem{REACL}    Astarita, G {\em Mass Transfer with Chemical Reactions}, Elsevier, New York {\bf 1967}.
\bibitem{EX1}      Stewart, M.E.; George, S.E.; Miller, R.L.; Paul, D.R. Polym.Eng and Sci {\bf 1993}, 33, 675.
\bibitem{EX3}      Ide, F.; Hasegawa, A. J.Appl.Poly.Sci. {\bf 1974}, 18, 963.
\bibitem{EX4}      Sundararaj, U.; Macosko, C.W. Macromolecules {\bf 1995}, 28, 2647.
\bibitem{EX5}      C.E. Scott, and C.W. Macosko, Polymer {\bf 35}, 5442 (1994).
\bibitem{EX6}      A description of recent experimental progress can be found in
		   Macromolecular Symposia: Polymer Blends {\bf 112} 141-175 (1996).
\bibitem{APPL}     Solc, C. (edt.) {\em Polymer Compatibility and Incompatibility, Principles and Practices},
                   Haarwood Academic, Chur, {\bf 1980}.\newline
                   Han, C.D. (edt.) {\em Polymer Blends and Composites in Multiphase Systems},
                   ACS, Washington D.C..\newline
                   Cahn, R.W.; Haasen, P.; Kramer, E.J.; {\em Materials Science and Technology, A Comprehensive Treatment},
                   Vol 12, VCH, Weinheim {\bf 1993}.
\bibitem{S1}       O'Shaughnessy, B.; Sawhey, U. Phys.Rev.Lett. {\bf 1996}, 76, 3444.
\bibitem{S2}       O'Shaughnessy, B.; Sawhey, U. Macromolecules {\bf 1996}, 29, 7230.
\bibitem{F1}       Fredrickson, G.H. Phys.Rev.Lett. {\bf 1996}, 76, 3440.
\bibitem{F2}       Fredrickson, G.H.; Milner, S.T. Macromolecules {\bf 1996}, 29, 7386.
\bibitem{BDYN}     Doi, M. Chem.Phys. {\bf 1975}, 9, 455.\newline
                   Doi, M. Chem.Phys. {\bf 1975}, 11, 107 and 115.\newline
                   deGennes, P.G. {\bf 1982}, J.Chem.Phys. 76, 3316.
\bibitem{ROUSE}    Rouse, P.E. J.Chem.Phys. {\bf 1953}, 21, 1272.
\bibitem{C1}       The crossover time $\tau$ can be estimated by equating the
		   number of copolymers produced per area in the initial regime (i.e. for times smaller than $\tau$)
		   $K_0\rho_0^2 \tau$ and the reduction of copolymers  per area in the depletion hole $\rho_0 \sqrt{D\tau}$
		   at time $\tau$.
\bibitem{EX2}      Guegan, P.; Macosko, C.W.; Ishizone, T.; Nakahama, S. Macromolecules {\bf 1994}, 27, 4993.
\bibitem{IDYN}     Durning, C.; O'Shaughnessy, B. J.Chem.Phys. {\bf 1988}, 88, 7117.
\bibitem{Albano}   see e.g.\ Gonzalez, A.P.; Pereyra, V.D.; Milchev, A.; Zgrablich, G. 
                   Phys.Rev.Lett. {\bf 1995}, 75, 3955.\newline
                   Satulovsky, J; Albano, E.V.; J.Chem.Phys. {\bf 1992}, 97, 9441.
\bibitem{BFM}      Carmesin, I.; Kremer, K. Macromolecules {\bf 1988}, 21, 2819.\newline
                   Deutsch, H.P.; Binder, K. J.Chem.Phys. {\bf 1991}, 94, 2294.
\bibitem{PAUL}     Paul, W.; Binder, K.; Heermann, D.; Kremer, K.  J.Chem.Phys. {\bf 1991}, 95, 7726.
\bibitem{NOTROUSE} M{\"u}ller, M.; Wittmer, J.P.; Cates, M.E. Phys.Rev.E {\bf 1996}, 53, 5063.
\bibitem{DYN}      M{\"u}ller, M.; Binder, K.; J.Phys. II France {\bf 1996}, 6, 187.
\bibitem{LIVE}     Rouault, Y.; Milchev, A. Phys.Rev.E {\bf 1995}, 51, 5903.
\bibitem{ACHIM}    Kopf, A.; Baschnagel, J.; Wittmer, J; Binder, K Macromolecules {\bf 1996}, 29, 1433.
\bibitem{INTER}    M{\"u}ller, M.; Binder, K.; Oed, W. Faraday Trans. {\bf 1995}, 91, 2369.
\bibitem{INTER2}   Schmid, F.; M{\"u}ller, M.  Macromolecules {\bf 1995}, 28, 8639.
\bibitem{TERNARY}  M{\"u}ller, M.; Schick, M. J.Chem.Phys. {\bf 1996}, 105, 8885.
\bibitem{WERNER}   Werner, A.; Schmid, F.; Binder, K.; M{\"u}ller, M. Macromolecules {\bf 1996}, 29, 8246.
\bibitem{MF}       Broseta, D.; Fredrickson, G.H. J.Chem.Phys. {\bf 1990}, 93, 2927.



\end{thebibliography}
\end{document}